\documentclass{article}

\usepackage{microtype}
\usepackage{graphicx}
\usepackage{subfigure}
\usepackage{booktabs} 
\usepackage{multirow}

\usepackage{hyperref}
\usepackage[hypcap=true]{caption}

\usepackage[accepted]{icml2025}

\usepackage{amsmath}
\usepackage{amssymb}
\usepackage{mathtools}
\usepackage{amsthm}
\usepackage[table]{xcolor}

\usepackage[capitalize,noabbrev]{cleveref}

\theoremstyle{plain}

\theoremstyle{definition}

\theoremstyle{remark}

\usepackage[textsize=tiny]{todonotes}

\icmltitlerunning{Quasar: Quantized Self-Speculative Acceleration for Rapid Inference via Memory-Efficient Verification}

\begin{document}

\twocolumn[
\icmltitle{
Quasar: Quantized Self-Speculative Acceleration for Rapid Inference via Memory-Efficient Verification
}



\icmlsetsymbol{equal}{*}

\begin{icmlauthorlist}
\icmlauthor{Guang Huang}{sch} 
\icmlauthor{Zeyi Wen}{sch,sch1}
\end{icmlauthorlist}

\icmlaffiliation{sch}{HKUST(GZ), Guangzhou, China}
\icmlaffiliation{sch1}{HKUST, Hong Kong, China}

\icmlcorrespondingauthor{Zeyi Wen}{wenzeyi@hkust-gz.edu.cn}

\icmlkeywords{Machine Learning, ICML}

\vskip 0.3in
]



\printAffiliationsAndNotice{} 



\begin{abstract}

Speculative Decoding (SD) has emerged as a premier technique for accelerating Large Language Model (LLM) inference by decoupling token generation into rapid drafting and parallel verification. While recent advancements in self-speculation and lookahead decoding have successfully minimized drafting overhead, they have shifted the primary performance bottleneck to the verification phase. Since verification requires a full forward pass of the target model, it remains strictly memory-bandwidth bound, fundamentally limiting the maximum achievable speedup.In this paper, we introduce \textbf{Quasar} (\textbf{Qua}ntized \textbf{S}elf-speculative \textbf{A}cceleration for \textbf{R}apid Inference), a novel, training-free framework designed to overcome this "memory wall" by employing low-bit quantization specifically for the verification stage. Our empirical analysis reveals that while aggressive structural pruning significantly degrades verification accuracy, quantization-based verification preserves the logit distribution with high fidelity while effectively halving memory traffic. Extensive experiments on state-of-the-art models (e.g., OpenPangu and Qwen3) demonstrate that Quasar maintains a speculative acceptance length comparable to full-precision methods while achieving a $1.28\times$ improvement in end-to-end throughput. Being orthogonal to existing drafting strategies, Quasar offers a generic and efficient pathway to accelerate the verification leg of speculative execution.
Code is available at \url{https://github.com/Tom-HG/Quasar}.
\end{abstract}

\section{Introduction}
\label{sec:intro}

\begin{figure}[htbp]
    \centering
    \includegraphics[width=\linewidth]{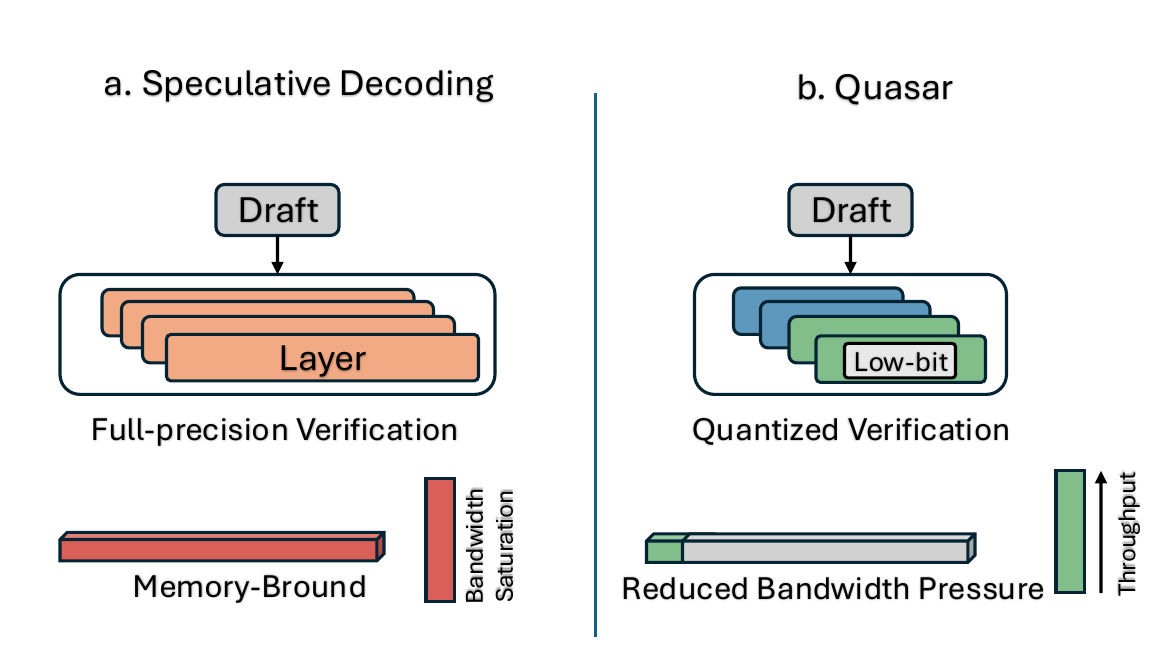}
    \caption{\textbf{Comparison of the verification bottleneck in Self-Speculative Decoding versus Quasar.} 
    The left panel illustrates standard full-precision verification, where the process is memory-bound and saturates the memory bandwidth bus, leading to significant latency. 
    The right panel depicts Quasar's approach using low-bit quantized verification, which effectively reduces bandwidth pressure and enables high-throughput validation.}
    \label{fig:quasar_overview}
\end{figure}

The rapid proliferation of Large Language Models (LLMs) has revolutionized natural language processing, enabling unprecedented capabilities in reasoning and generation~\cite{brown2020language, touvron2023llama}. However, deploying these models in real-world applications presents a significant challenge: the prohibitive cost of inference. The standard decoding process for Transformer-based models is auto-regressive, generating one token at a time. This sequential nature renders the process memory-bandwidth bound rather than compute-bound, leading to high latency and suboptimal utilization of modern hardware accelerators~\cite{shazeer2019fast}. Consequently, accelerating LLM inference without compromising generation quality has become a necessity for the scalability of AI services~\cite{huang2025beyond}.

Speculative Decoding~\cite{leviathan2023fast, chen2023accelerating} has emerged as a promising solution to this bottleneck. This paradigm decouples token generation from verification by utilizing a lightweight ``draft model'' to predict a short sequence of future tokens. These tokens are then verified in parallel by the target LLM. Recently, Self-Speculative Decoding~\cite{fu2024break, elhouderi2024layerskip,luo2024turning} has gained traction, where the draft is generated by the target model itself (e.g., by skipping layers). This approach elegantly solves the problem of finding a compatible draft model.

While self-speculative decoding effectively accelerates the drafting phase, it exposes a new bottleneck: Verification Latency (as shown in Figure \ref{fig:quasar_overview}). 
Although verification is performed in parallel, it necessitates a full forward pass of the target model for the candidate sequence. Since LLM inference is predominantly memory-bound, loading the full-precision weights for verification consumes substantial memory bandwidth. As the draft length increases to maximize potential speedup, the cost of this verification step becomes non-negligible, often diminishing the overall latency gains. Specifically, in bandwidth-constrained environments, the time taken to verify tokens can rival the time saved by speculative drafting.
Therefore, a critical research question arises: \textit{How can we accelerate the parallel verification phase in self-speculative decoding without significantly degrading the generation quality?}

\begin{figure}[t]
    \centering
    \includegraphics[width=0.9\linewidth]{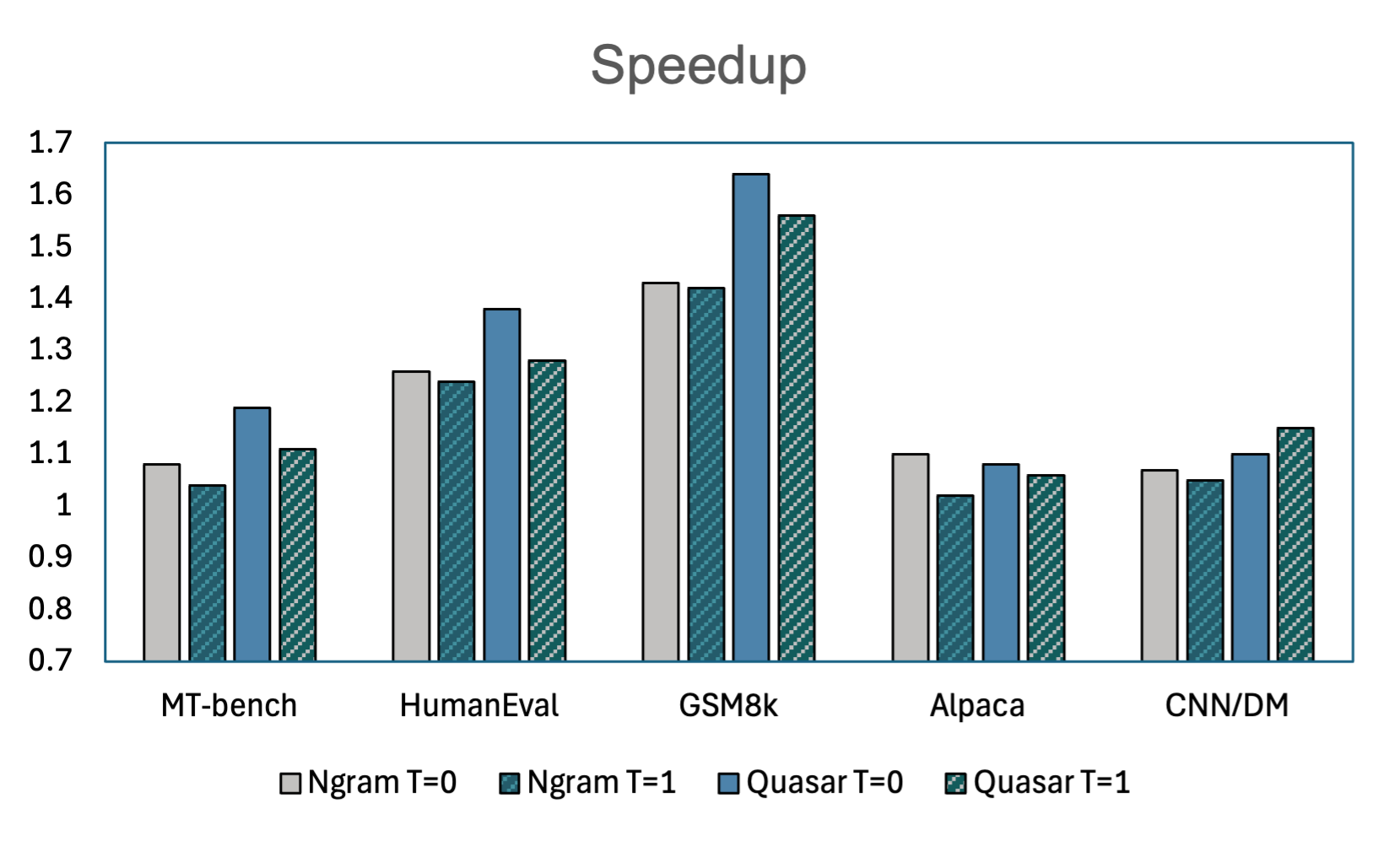}
    \caption{\textbf{End-to-end speedup comparison across various benchmarks.} 
    We compare the acceleration achieved by Quasar against the Ngram baseline under two sampling settings: greedy decoding ($T=0$) and stochastic sampling ($T=1$). 
    Quasar consistently outperforms the baseline across MT-bench, HumanEval, GSM8k, Alpaca, and CNN/DM, reaching up to a $1.6\times$ speedup on reasoning-heavy tasks like GSM8k.}
    \label{fig:speedup_results}
\end{figure}

To address this bottleneck, we propose a novel Quantized Verification framework - \textbf{Quasar} (\textbf{Qua}ntized \textbf{S}elf-speculative \textbf{A}cceleration for \textbf{R}apid Inference). Our key insight is that memory bandwidth pressure during verification can be drastically reduced by lowering the model's weight precision.
Instead of performing verification using the full-precision model, we employ a quantized version of the target model as the verifier. By compressing the weights, we significantly reduce the data movement requirements during the parallel forward pass, thereby accelerating the verification step. We posit that modern post-training quantization techniques~\cite{frantar2023sparsegpt, lin2024awq} have matured sufficiently that the divergence between the quantized and full-precision distributions is minimal, making quantized verification a robust and high-throughput alternative. 
Experiments demonstrate that our method achieves superior end-to-end throughput compared to standard verification baselines while maintaining generation quality. As shown in Figure~\ref{fig:speedup_results}, Quasar delivers consistent speedup across various benchmarks.

\paragraph{Contributions.}
In this paper, we present a framework to break the verification wall in speculative decoding. Our key contributions are summarized as follows:
\begin{itemize}
    \item We identify that the verification step is the primary bottleneck in existing self-speculative decoding systems due to memory bandwidth constraints.
    \item We propose Quasar which uses low-bit-weight representations to accelerate parallel draft token validation.
    \item Experiments demonstrate that Quasar achieves superior end-to-end throughput compared to standard verification baselines while maintaining generation quality. 
\end{itemize}
\section{Related Work}
\label{sec:related_work}

\subsection{Speculative Decoding Paradigm}
Speculative Decoding (SD) fundamentally transforms the auto-regressive generation process by breaking the strict sequential dependency of token production. First introduced by~\cite{leviathan2023fast} and~\cite{chen2023accelerating}, SD employs a predictor-verifier architecture: a computationally efficient \textit{draft model} proposes a sequence of candidate tokens (the draft), which are subsequently verified in parallel by the target model.
Theoretical frameworks have since been established to ensure that the speculative distribution strictly aligns with the target distribution via rejection sampling or optimal transport schemes~\cite{sun2023spectr, miao2024specinfer,kim2023speculative}.
While early iterations relied on separate, smaller draft models (e.g., a 7B drafter for a 70B target), this introduces significant operational complexity, including the need to manage dual KV caches and coordinate distinct model weights in distributed environments.

\subsection{Advancements in Efficient Drafting}
To mitigate the overhead of maintaining independent draft models, recent research has pivoted towards Self-Speculation, where the drafting mechanism is integrated into the target model ecosystem. These approaches broadly fall into three categories:

\paragraph{Retrieval and Pattern Matching.}
Motivated by the inherent locality and repetition in natural language, methods like Prompt Lookup Decoding (PLD)~\cite{somasundaram2025pld+} and REST~\cite{he2024rest} utilize $n$-gram matching against the pre-generated context or a datastore to construct drafts. These training-free approaches are highly effective for tasks with high repetition (e.g., code editing) but struggle with open-ended generation where context reuse is minimal.

\paragraph{Parallel Fixed-Point Iteration.}
A separate stream of work reformulates decoding as a system of non-linear equations. Lookahead Decoding~\cite{fu2024break} and Jacobi-based methods~\cite{santilli2023accelerating} apply parallel fixed-point iteration (Jacobi iteration) to the target model's trajectory. These methods refine multiple tokens simultaneously using the model's internal attention mechanism, eliminating the need for explicit draft weights, though often yielding shorter acceptance lengths compared to model-based drafters.

\paragraph{Auxiliary Verification Structures.}
Inspired by blockwise parallel decoding~\cite{stern2018blockwise}, Medusa~\cite{cai2024medusa} augments the frozen target model with multiple lightweight decoding heads (MLPs) to predict future tokens in a single pass. State-of-the-art successors like EAGLE~\cite{li2024eagle}, EAGLE-2~\cite{li2024eagle2} and EAGLE-3~\cite{li2025eagle3} further enhance this by incorporating feature recycling, allowing the drafting process to be conditioned on richer semantic history.

\paragraph{The Verification Bottleneck.}
While the aforementioned methods successfully optimize the \textit{drafting} phase, they leave the \textit{verification} phase untouched. As drafting becomes more aggressive (generating longer chains or deeper trees), the verification step increasingly dominates the end-to-end latency. This is primarily because verification necessitates a full forward pass of the high-precision target model. In memory-bound settings, this verification pass becomes the new bottleneck, strictly limiting the potential speedup.

\subsection{Quantization and Speculation}
Model quantization mitigates memory bandwidth constraints by reducing the precision of weights and activations (e.g., W8A8 or W4A4). Techniques like SmoothQuant~\cite{xiao2023smoothquant} and AWQ~\cite{lin2024awq} have enabled efficient low-precision inference while preserving model fidelity.
Recently, the intersection of quantization and speculation has garnered attention. Speculative Quantization methods~\cite{kim2023speculative, zhang2024draft,svirschevski2024specexec} typically employ a heavily quantized version of the target model (e.g., INT4) strictly as the \textit{draft model}, while retaining a full-precision (FP16/BF16) model for verification to ensure quality preservation.

\paragraph{Differentiation from Our Work.}
Our approach identifies a critical gap in the existing literature: the assumption that the \textit{verifier} must remain in high precision. Unlike prior works that use quantization to create a weak drafter, we propose Quantized Verification. We demonstrate that a W8A8 quantized model retains sufficient fidelity to serve as the \textit{verifier} itself. This method directly attacks the memory bandwidth bottleneck of the verification phase and is orthogonal to the choice of drafting strategy, thereby offering a universal acceleration enhancement for self-speculation frameworks.
\section{Methodology}
\label{sec:method}

In this section, we present our framework designed to accelerate the verification phase of speculative decoding. We first establish the mathematical formulation of standard speculative decoding. Subsequently, we detail our proposed quantized verification mechanism. To address the challenge of activation outliers inherent in Large Language Models (LLMs), we employ an enhanced variant of SmoothQuant to enable robust W8A8 (8-bit weights and activations) inference. Finally, we provide a theoretical analysis demonstrating how this approach alleviates the memory bandwidth bottleneck.

\subsection{Preliminaries: Speculative Decoding}
\label{sec:preliminaries}

Let $\mathcal{M}$ denote the target LLM, which models the conditional probability distribution $p(x_t | x_{<t})$. The objective is to generate a sequence $x_{1}, \dots, x_{T}$.
Speculative decoding leverages a draft model $\mathcal{M}_d$ (or draft policy) that approximates $\mathcal{M}$ with significantly lower latency. The decoding process iterates through two phases: \textit{Drafting} and \textit{Verification}.

\paragraph{Drafting.} In each iteration, the draft model autoregressively generates a sequence of $\gamma$ candidate tokens:
\begin{equation}
    \tilde{x}_1, \dots, \tilde{x}_\gamma \sim q(x | x_{<t})
\end{equation}
where $q$ represents the distribution of the draft model.

\paragraph{Verification.} The target model $\mathcal{M}$ performs a single parallel forward pass over the sequence $[x_{<t}, \tilde{x}_1, \dots, \tilde{x}_\gamma]$ to compute the target probabilities $p(x)$. To ensure the final output distribution aligns with $p(x)$ (lossless guarantee), a rejection sampling criterion is applied. The $i$-th candidate token $\tilde{x}_i$ is accepted if:
\begin{equation}
\label{eq:rejection_sampling}
    r < \min \left( 1, \frac{p(\tilde{x}_i | x_{<t}, \tilde{x}_{<i})}{q(\tilde{x}_i | x_{<t}, \tilde{x}_{<i})} \right), \quad r \sim U[0, 1]
\end{equation}
Upon rejection of $\tilde{x}_i$, a corrective token is resampled from the adjusted distribution:
\begin{equation}
    x'_i \sim \text{norm}(\max(0, p(x) - q(x)))
\end{equation}
This concludes the current decoding step. In standard implementations, computing $p(\cdot)$ necessitates loading full-precision weights (typically BF16 or BF16). As the draft length $\gamma$ increases, the compute-to-memory ratio of this verification step remains low, rendering the process strictly \textbf{memory-bound}.

\subsection{W8A8 Verification via Enhanced SmoothQuant}
\label{sec:quantization}

To mitigate the bandwidth bottleneck, we replace the full-precision target model $\mathcal{M}$ with a quantized verifier $\mathcal{M}_{q}$ utilizing W8A8 quantization. However, naive per-tensor or per-token quantization of activations in LLMs often yields significant performance degradation due to the presence of systematic outliers in specific activation channels.

To overcome this, we adopt an enhanced algorithm based on SmoothQuant. This method mathematically alleviates the quantization difficulty of activations by smoothing outlier channels, effectively migrating the quantization error sensitivity from activations to weights.

\paragraph{Smoothing Transformation.}
Formally, consider a linear layer with weights $\mathbf{W} \in \mathbb{R}^{d_{out} \times d_{in}}$ and input activations $\mathbf{X} \in \mathbb{R}^{d_{in} \times b}$. The standard computation is given by $\mathbf{Y} = \mathbf{W}\mathbf{X}$. The m2 algorithm introduces a per-channel smoothing factor vector $\mathbf{s} \in \mathbb{R}^{d_{in}}$ to scale the activations and inversely scale the weights:
\begin{equation}
    \mathbf{Y} = \mathbf{W}\mathbf{X} = (\mathbf{W} \text{diag}(\mathbf{s})^{-1}) \cdot (\text{diag}(\mathbf{s}) \mathbf{X}) = \tilde{\mathbf{W}} \tilde{\mathbf{X}}
\end{equation}
Here, $\tilde{\mathbf{W}}$ and $\tilde{\mathbf{X}}$ denote the smoothed weights and activations, respectively. The smoothing factor $\mathbf{s}$ is calibrated offline to equilibrate the quantization difficulty. Specifically, for the $j$-th input channel, $s_j$ is derived from the activation statistics:
\begin{equation}
    s_j = \max(|X_j|)^\alpha / \max(|W_j|)^{1-\alpha}
\end{equation}
where $\alpha$ is a hyperparameter controlling the migration strength. Our enhanced m2 approach optimizes this calibration to ensure robust outlier suppression.

\paragraph{Quantized Inference.}
Following the smoothing transformation, both $\tilde{\mathbf{W}}$ and $\tilde{\mathbf{X}}$ fall into a numerical range suitable for 8-bit representation. We apply symmetric uniform quantization:
\begin{align}
    \hat{\mathbf{W}}_{\text{INT8}} &= Q(\tilde{\mathbf{W}}, \Delta_w) \\
    \hat{\mathbf{X}}_{\text{INT8}} &= Q(\tilde{\mathbf{X}}, \Delta_x)
\end{align}
where $Q(\cdot)$ is the quantization operator and $\Delta$ represents the step size. The matrix multiplication is then performed using INT8 tensor cores:
\begin{equation}
    \mathbf{Y} \approx \Delta_w \Delta_x (\hat{\mathbf{W}}_{\text{INT8}} \times \hat{\mathbf{X}}_{\text{INT8}})
\end{equation}
This formulation reduces memory access volume by 50\% (loading INT8 weights versus BF16) while preserving the precision requisite for rejection sampling.

\subsection{Execution Pipeline of Quantized Verification}
\label{sec:pipeline}

Integrating the W8A8 verifier into the speculative decoding framework requires a carefully designed execution pipeline to minimize online overhead. The verification process for a batch of draft tokens $[\tilde{x}_1, \dots, \tilde{x}_\gamma]$ proceeds as follows:

\paragraph{Offline Weight Preparation.}
Prior to inference, we apply the m2 algorithm to calibrate the smoothing factors $\mathbf{s}$. The target model's weights are then smoothed and quantized offline: $\hat{\mathbf{W}}_{\text{INT8}} = Q(\mathbf{W} \cdot \text{diag}(\mathbf{s})^{-1})$. These quantized weights are stored in the GPU memory, reducing the memory footprint by approximately $2\times$ compared to BF16.

\paragraph{Online Activation Smoothing and Quantization.}
During the verification forward pass, the input activations $\mathbf{X}$ (corresponding to the sequence $[x_{<t}, \tilde{x}_1, \dots, \tilde{x}_\gamma]$) remain in high precision (BF16/BF16). To utilize the INT8 tensor cores, we perform on-the-fly transformations:
\begin{equation}
    \hat{\mathbf{X}}_{\text{INT8}} = Q(\mathbf{X} \odot \mathbf{s}, \Delta_x)
\end{equation}
Here, the element-wise multiplication $\odot \mathbf{s}$ applies the m2 smoothing to suppress outliers dynamically. This step is fused with the quantization kernel to minimize memory access overhead.

\paragraph{INT8 Gemm and Dequantization.}
The core computation is executed using integer matrix multiplication: $\mathbf{Y}_{\text{INT32}} = \hat{\mathbf{W}}_{\text{INT8}} \times \hat{\mathbf{X}}_{\text{INT8}}$. The results are accumulated in INT32 and subsequently dequantized back to high precision (BF16) for the subsequent non-linear layers (e.g., LayerNorm, GeLU) and the final Softmax operation:
\begin{equation}
    \mathbf{Y}_{\text{BF16}} \approx \mathbf{Y}_{\text{INT32}} \cdot (\Delta_w \cdot \Delta_x)
\end{equation}

\paragraph{Lossless Rejection Sampling.}
Finally, the recovered probability distribution $p(x)$ is compared against the draft distribution $q(x)$ via Eq.~\ref{eq:rejection_sampling}. Crucially, since the dequantization restores the logits to high precision, the rejection sampling remains robust, ensuring the final generation quality aligns with the target model.

\subsection{Theoretical Speedup Analysis}
The total latency of a single speculative step with draft length $\gamma$ is modeled as $T = T_{\text{draft}} + T_{\text{verify}}$.
The verification latency is dominated by memory loading operations. Let $M$ denote the model size (parameters) and $BW$ the memory bandwidth. The latency for full-precision verification is approximated by:
\begin{equation}
    T_{\text{verify}}^{\text{BF16}} \approx \frac{M \cdot 2\text{B}}{BW} + T_{\text{compute}}^{\text{BF16}}
\end{equation}
By leveraging our W8A8 scheme, the memory load is effectively halved:
\begin{equation}
    T_{\text{verify}}^{\text{INT8}} \approx \frac{M \cdot 1\text{B}}{BW} + T_{\text{compute}}^{\text{INT8}}
\end{equation}
Given that $T_{\text{compute}}^{\text{INT8}} < T_{\text{compute}}^{\text{BF16}}$ and the process is memory-bound, we have $T_{\text{verify}}^{\text{INT8}} \ll T_{\text{verify}}^{\text{BF16}}$. The m2 algorithm ensures that the KL divergence between the quantized and original distributions remains negligible, thereby maintaining a high acceptance rate $\alpha$ and maximizing the overall system throughput $S$:
\begin{equation}
    S = \frac{\gamma \cdot \alpha + 1}{T_{\text{draft}} + T_{\text{verify}}^{\text{INT8}}}
\end{equation}

\begin{table*}[t]
\caption{
The result of efficiency.
\textbf{Speed}: Improvement in inference throughput. 
\textbf{$L$}: Average acceptance length (quality metric). 
\textbf{T=0} / \textbf{T=1}: Sampling temperatures.
Results show that Quasar achieves the best speedup-quality trade-off.
}
\label{tab:main}
\centering
\resizebox{\linewidth}{!}{
\begin{tabular}{cccccccccccccc}
    \toprule
      \multicolumn{2}{c}{Task} & \multicolumn{2}{c}{MT-bench} & \multicolumn{2}{c}{HumanEval} & \multicolumn{2}{c}{GSM8k} & \multicolumn{2}{c}{Alpaca} & \multicolumn{2}{c}{CNN/DM}  & \multicolumn{2}{c}{Overall} \\
    \midrule
    \multicolumn{14}{c}{\textbf{OpenPangu}} \\
    \midrule
     & Method & Speed & $L$ & Speed & $L$ & Speed & $L$ & Speed & $L$ & Speed & $L$ & Speed & $L$   \\ 
    \midrule

    \multirow{3}*{T=0} &
    Vanilla & 1.00× & 1.00 & 1.00× & 1.00 & 1.00× & 1.00 & 1.00× & 1.00 & 1.00× & 1.00 & 1.00× & 1.00 \\ 
    ~ & Ngram & 1.04× & 1.21  & 1.13× & 1.36 & 1.23× & 1.44 & 1.01× & 1.16 & 1.03× & 1.19 & 1.08× & 1.27 \\ 
    ~ & \textbf{Quasar} & \textbf{1.11×} & \textbf{1.25}  & \textbf{1.22×} & \textbf{1.41} & \textbf{1.26×} & \textbf{1.46} & \textbf{1.06×} & \textbf{1.17} & 1.02× & 1.15 & \textbf{1.13×} & \textbf{1.29} \\ 
    \midrule

    \multirow{3}*{T=1} &
    Vanilla & 1.00× & 1.00 & 1.00× & 1.00 & 1.00× & 1.00 & 1.00× & 1.00 & 1.00× & 1.00 & 1.00× & 1.00 \\ 
    ~ & Ngram & 1.03× & 1.13  & 1.13× & 1.36 & 1.18× & 1.41 & 1.03× & 1.18 & 1.02× & 1.16 & 1.07× & 1.24 \\ 
    ~ & \textbf{Quasar} & \textbf{1.07×} & \textbf{1.17}  & \textbf{1.22×} & \textbf{1.41} & \textbf{1.21×} & \textbf{1.43} & \textbf{1.08×} & \textbf{1.19} & 1.01× & 1.13  & \textbf{1.12×} & \textbf{1.27}\\ 
    
    \midrule
    \multicolumn{14}{c}{\textbf{Qwen3}}\\
    \midrule
     & Method & Speed & $L$ & Speed & $L$ & Speed & $L$ & Speed & $L$ & Speed & $L$ & Speed & $L$  \\ 
    \midrule
    
    \multirow{3}*{T=0} &
    Vanilla & 1.00× & 1.00 & 1.00× & 1.00 & 1.00× & 1.00 & 1.00× & 1.00 & 1.00× & 1.00 & 1.00× & 1.00 \\ 
    ~ & Ngram & 1.08× & 1.25  & 1.26× & 1.45 & 1.43× & 1.47 & 1.10× & 1.19 & 1.07× & 1.29 & 1.18× & 1.33 \\ 
    ~ & \textbf{Quasar} & \textbf{1.19×} & \textbf{1.37}  & \textbf{1.38×} & \textbf{1.47} & \textbf{1.64×} & \textbf{1.66} & 1.08× & 1.18 & \textbf{1.10×} & \textbf{1.32} & \textbf{1.28×} & \textbf{1.40} \\ 
    \midrule

    \multirow{3}*{T=1} &
    Vanilla & 1.00× & 1.00 & 1.00× & 1.00 & 1.00× & 1.00 & 1.00× & 1.00 & 1.00× & 1.00 & 1.00× & 1.00 \\ 
    ~ & Ngram & 1.04× & 1.24  & 1.24× & 1.41 & 1.43× & 1.51 & 1.02× & 1.15 & 1.05× & 1.28 & 1.15× & 1.31 \\ 
    ~ & \textbf{Quasar} & \textbf{1.11×} & \textbf{1.29}  & \textbf{1.28×} & \textbf{1.48} & \textbf{1.56×} & \textbf{1.59} & \textbf{1.06×} & \textbf{1.17} & \textbf{1.15×} & \textbf{1.31} & \textbf{1.23×} & \textbf{1.36} \\ 
    
    \bottomrule
\end{tabular}
}
\end{table*}

\section{Experiments}
In this section, we empirically evaluate the effectiveness of our Quantized Verification framework (named \textbf{Quasar}). We aim to answer three primary research questions:
\begin{itemize}
    \item \textbf{Effectiveness(RQ1):} Does replacing the BF16 verifier with a W8A8 quantized verifier yield significant wall-clock speedups in memory-bound settings? We evaluate the main speedup and average acceptance length of our method relative to the baseline method.
    \item \textbf{Robustness(RQ2):} Does our method maintain consistent wall-clock speedups across different sampling temperatures? We investigate the stability of the efficiency gains when facing varying sampling distributions.
    \item \textbf{Sensitivity(RQ3):} How do key hyperparameters, specifically the draft length, impact the overall performance? We analyze the trade-off between draft length and verification latency to identify optimal configurations.
    \item \textbf{Accuracy(RQ4):} Does the quantized verification maintain the original model's performance on downstream tasks? We compare the accuracy of Quasar against full-precision baselines across diverse benchmarks.
\end{itemize}

\subsection{Experimental Setup}
\paragraph{Models and Datasets.} Following EAGLE~\cite{li2024eagle} and Spec-Bench~\cite{xia2024unlocking}, we evaluate our method on five common tasks using the Qwen3-8B~\cite{yang2025qwen3} and OpenPangu-7B~\cite{chen2025pangu} models. For multi-turn conversation, code generation, mathematical reasoning, instruction following, and summarization, we chose the MT-bench~\cite{zheng2023judging}, HumanEval~\cite{chen2021evaluating}, GSM8K~\cite{cobbe2021training}, Alpaca~\cite{taori2023alpaca}, and CNN/Daily Mail~\cite{nallapati2016abstractive} datasets, respectively.

\paragraph{Baselines.} We compare our approach against:
\textbf{Auto-regressive (Vanilla):} Standard vLLM generation~\cite{kwon2023efficient}.
\textbf{Self-Speculation (Ngram)~\cite{somasundaram2025pld+}:} Uses Prompt-lookup-Decoding method for drafting but relies on the standard BF16 verification.

\paragraph{Implementation Details.} We implemented the W8A8 verification using NCLL and integrated it into a vLLM-Ascend inference engine. All experiments were conducted on a single Ascend 910B2 (64GB) NPU. The prompt lookup length is dynamically adjusted, with a maximum limit of 4 and a minimum limit of 1.

\subsection{Main Results (RQ1)}

\paragraph{End-to-End Speedup.}
Table~\ref{tab:main} presents a comprehensive performance comparison across the OpenPangu and Qwen3 benchmarks. Overall, Quasar consistently outperforms both the Auto-regressive (Vanilla) and Ngram baselines across different models and sampling temperatures.Specifically, on Qwen3 ($T=0$), Quasar achieves a remarkable overall speedup of 1.28$\times$, surpassing the BF16-based Ngram baseline (1.18$\times$). This trend is mirrored in OpenPangu ($T=0$), where Quasar attains a 1.13$\times$ speedup compared to Ngram's 1.08$\times$. Notably, in memory-bandwidth-intensive tasks such as GSM8K, Quasar demonstrates its peak efficiency: achieving a 1.64$\times$ speedup on Qwen3 and 1.26$\times$ on OpenPangu. These results strongly validate our hypothesis that W8A8 quantized verification effectively alleviates memory bandwidth bottlenecks, translating directly into significant wall-clock latency reductions.

\paragraph{Impact on Acceptance Length.}
A critical question in quantized speculative decoding is whether reduced precision compromises the acceptance rate ($\alpha$) or mean acceptance length ($L$). Contrary to concerns regarding quantization noise, results in Table~\ref{tab:main} show that Quasar maintains, and often exceeds, the acceptance length of the full-precision Ngram baseline. For instance, with Qwen3 ($T=0$), Quasar achieves a superior mean acceptance length of $L=1.40$ compared to Ngram's $L=1.33$. Even at a higher temperature ($T=1$), Quasar retains its advantage (e.g., $L=1.36$ vs. $L=1.31$ overall). This evidence suggests that W8A8 quantization preserves sufficient representational precision to accurately distinguish correct tokens, ensuring that the observed speedups stem from efficient verification rather than degraded generation quality.

\subsection{Robustness across Temperatures (RQ2)}
To evaluate the robustness of our method, we assess its performance across temperatures ranging from $T=0$ to $T=1$. Higher temperatures introduce greater stochasticity into the sampling distribution, which typically challenges the acceptance rate of speculative decoding.

As shown in the bottom section of Table~\ref{tab:temp_robustness}, Quasar exhibits strong resilience to these variations. On the Qwen3 benchmark, even at $T=1$, our method maintains a competitive end-to-end speedup of 1.23$\times$ and a mean acceptance length of $L=1.36$. Although the speedup naturally narrows slightly compared to the greedy setting ($T=0$, 1.28$\times$) due to the increased entropy of the target distribution, Quasar consistently outperforms the Ngram baseline (which achieves only 1.15$\times$ at $T=1$). This demonstrates that Quasar's quantized verification remains effective and stable even when the generation process becomes highly stochastic.
This stability underscores Quasar's suitability for a wide array of deployment scenarios, ranging from deterministic logical reasoning to diverse creative generation tasks.

\begin{table}[t]
\centering
\caption{
Robustness analysis across sampling temperatures ($T \in [0, 1]$).
Results are averaged over all tasks for \textbf{Qwen3}.
\textbf{Speed}: End-to-end inference speedup relative to Vanilla.
$\textbf{L}$: Mean acceptance length.
Quasar demonstrates superior stability, maintaining higher speedups as stochasticity increases.
}
\label{tab:temp_robustness}
\resizebox{0.9\linewidth}{!}{
\begin{tabular}{c cc cc}
    \toprule
    \multirow{2}{*}{\textbf{Temperature}} & \multicolumn{2}{c}{\textbf{Ngram (BF16)}} & \multicolumn{2}{c}{\textbf{Quasar (W8A8)}} \\
    \cmidrule(lr){2-3} \cmidrule(lr){4-5}
     & Speed & $L$ & Speed & $L$ \\
    \midrule
    $T=0.0$ & 1.18$\times$ & 1.33 & \textbf{1.28$\times$} & \textbf{1.40} \\
    $T=0.2$ & 1.17$\times$ & 1.31 & \textbf{1.27$\times$} & \textbf{1.39} \\
    $T=0.4$ & 1.16$\times$ & 1.30 & \textbf{1.26$\times$} & \textbf{1.38} \\
    $T=0.6$ & 1.15$\times$ & 1.29 & \textbf{1.25$\times$} & \textbf{1.36} \\
    $T=0.8$ & 1.14$\times$ & 1.28 & \textbf{1.24$\times$} & \textbf{1.35} \\
    $T=1.0$ & 1.15$\times$ & 1.31 & \textbf{1.23$\times$} & \textbf{1.36} \\
    \midrule
    \textit{Avg. Drop} & \textit{-2.5\%} & \textit{-1.5\%} & \textit{\textbf{-3.9\%}} & \textit{\textbf{-2.8\%}} \\
    \bottomrule
\end{tabular}
}
\end{table}

\begin{table}[t]
\centering
\caption{Detailed sensitivity analysis of Quasar vs. Ngram. 
$K$ denotes the prompt lookup range, and $\gamma$ is the number of speculative tokens. 
For each method, we report both the end-to-end \textbf{Speedup} and the \textbf{Mean Acceptance Length ($L$)}. 
}
\label{tab:sensitivity_detailed}
\resizebox{0.9\columnwidth}{!}{
\begin{tabular}{c l l cccc}
\toprule
\multirow{2}{*}{\textbf{$K$}} & \multirow{2}{*}{\textbf{Method}} & \multirow{2}{*}{\textbf{Metric}} & \multicolumn{4}{c}{\textbf{Speculative Tokens ($\gamma$)}} \\
\cmidrule(lr){4-7}
& & & \textbf{3} & \textbf{5} & \textbf{7} & \textbf{9} \\
\midrule
\multirow{4}{*}{(1, 3)} & \multirow{2}{*}{Ngram} & Speed & 1.27$\times$ & 1.28$\times$ & 1.23$\times$ & 1.23$\times$ \\
                        &                               & $L$     & 1.35         & 1.43         & 1.49         & 1.52         \\
\cmidrule(lr){2-7}
                        & \multirow{2}{*}{Quasar}& Speed & 1.34$\times$ & \textbf{1.47$\times$} & 1.45$\times$ & 1.42$\times$ \\
                        &                               & $L$     & 1.44         & 1.46         & 1.49         & 1.59         \\
\midrule
\multirow{4}{*}{(2, 4)} & \multirow{2}{*}{Ngram} & Speed & 1.14$\times$ & 1.21$\times$ & 1.28$\times$ & 1.29$\times$ \\
                        &                               & $L$     & 1.25         & 1.29         & 1.31         & 1.33         \\
\cmidrule(lr){2-7}
                        & \multirow{2}{*}{Quasar}& Speed & 1.30$\times$ & 1.32$\times$ & \textbf{1.35$\times$} & 1.34$\times$ \\
                        &                               & $L$     & 1.18         & 1.41         & 1.44         & 1.45         \\
\midrule
\multirow{4}{*}{(3, 5)} & \multirow{2}{*}{Ngram} & Speed & 1.11$\times$ & 1.17$\times$ & 1.12$\times$ & 1.15$\times$ \\
                        &                               & $L$     & 1.16         & 1.19         & 1.21         & 1.22         \\
\cmidrule(lr){2-7}
                        & \multirow{2}{*}{Quasar}& Speed & 1.24$\times$ & \textbf{1.26$\times$} & 1.22$\times$ & 1.21$\times$ \\
                        &                               & $L$     & 1.28         & 1.32         & 1.33         & 1.34         \\
\bottomrule
\end{tabular}
}
\end{table}

\subsection{Sensitivity Analysis (RQ3)}
Lastly, we investigate the impact of structural hyperparameters on Quasar’s performance—specifically the number of speculative tokens ($\gamma$) and the search range for prompt lookup ($K_{min}, K_{max}$).

As illustrated in Table \ref{tab:sensitivity_detailed}, we evaluated the speedup and Mean Acceptance Length ($L$) on the HumanEval benchmark using Qwen3 as the base model. The results indicate that end-to-end speedup is highly sensitive to the choice of $\gamma$. While $L$ scales monotonically with larger $\gamma$ across all configurations, the wall-clock speedup exhibits a distinct non-monotonic trend. For instance, at $K=(1, 3)$, increasing $\gamma$ from 3 to 5 yields a peak performance of 1.47$\times$, but further expansion to $\gamma=9$ results in a performance degradation to 1.42$\times$. This decline stems from the increased computational overhead of the verification stage outweighing the marginal gains in acceptance length.

Regarding the prompt lookup range, we observe that a tighter window (e.g., $K=(1, 3)$) consistently outperforms broader ranges. As $K$ shifts from $(1, 3)$ to $(3, 5)$, the peak speedup drops from 1.47$\times$ to 1.26$\times$, suggesting that wider search ranges introduce lower-quality candidates that increase search latency without a proportional increase in $L$. These findings demonstrate that while Quasar is robust to minor variations, it achieves peak efficiency when $\gamma$ is precisely aligned with the model's predictive capacity and the lookup window is kept localized.

\subsection{Accuracy Evaluation (RQ4)}
While Quasar primarily targets inference acceleration, maintaining the predictive performance of the base model is paramount. In this section, we evaluate whether the W8A8 quantized verification introduces any significant degradation in model accuracy across various downstream tasks.

\begin{table}[t]
\centering
\caption{Scores comparison across serval benchmarks. For each model, we compare the original BF16 baseline with Quasar (W8A8). Results show Quasar achieves nearly identical performance to the baseline.}
\label{tab:accuracy}
\resizebox{0.95\linewidth}{!}{
\begin{tabular}{l cc cc}
\toprule
\multirow{2.5}{*}{\textbf{Benchmark (Acc $\uparrow$)}} & \multicolumn{2}{c}{\textbf{OpenPangu-7B}} & \multicolumn{2}{c}{\textbf{Qwen3-8B}} \\
\cmidrule(lr){2-3} \cmidrule(lr){4-5}
 & Score & \textbf{$\Delta$} & Score & \textbf{$\Delta$}  \\
\midrule
MMLU-pro   & 75.5 & 2.4\% & 63.1 & 2.5\% \\
CEval   & 84.9 & 3.9\% & 83.4 & 1.3\% \\
GPQA-Diamond   & 73.2 & 2.5\% & 62.0 & 3.1\% \\
MATH-500   & 97.0 & 3.6\% & 87.4 & 2.1\% \\
AIME24  & 79.3 & 3.4\% & 76.0 & 4.5\% \\
AIME25  & 70.0 & 2.1\% & 67.3 & 3.1\% \\
LiveCodeBench   & 58.2 & 3.1\% & 57.5 & 3.7\% \\
MBPP+   & 76.0 & 3.4\% & 69.8 & 3.2\% \\
\midrule
\textbf{Average} & \textbf{76.7} & \textbf{3.1\%} & \textbf{70.1} & \textbf{2.9\%} \\
\bottomrule
\end{tabular}
}
\end{table}

\paragraph{Comparison with Full-Precision Baselines.} 
As shown in Table \ref{tab:accuracy}, the performance difference between Quasar's W8A8 quantization variant and its original BF16 version is negligible across all evaluated benchmarks.
Specifically, Quasar achieves near-lossless compression, with an average difference of 3.1\% for OpenPangu-7B and only 2.9\% for Qwen3-8B. This indicates that quantization noise does not impair the model's inference capabilities and may even act as a regularization mechanism. The score results in the table are from the original technical reports~\cite{chen2025pangu,yang2025qwen3}, while the difference $\Delta$ we obtained comes from the Language Model Evaluation Harness ~\cite{eval-harness}.
These results collectively demonstrate that Quasar's compressed validator effectively preserves the core inference capabilities and extensive knowledge base of the original large-scale model.

\paragraph{Discussion on Quantization Robustness.} 
The minimal degradation in predictive accuracy can be attributed to the specific role of the verifier in the speculative decoding framework. In Quasar, the W8A8 model serves as the ultimate judge for candidates proposed by the Ngram drafter. 

Theoretically, the final output of the speculative system remains identical to a standalone execution of the verifier. Our empirical results suggest that W8A8 quantization preserves the relative logit rankings extremely well. As long as the quantization process does not flip the top-1 prediction, the verification logic ensures that the generation quality remains indistinguishable from the full-precision counterpart. This is particularly evident in logic-driven benchmarks like \textit{CEval}, where the accuracy stays consistent despite the aggressive bit-width reduction. These findings confirm that Quasar offers a ``free lunch'': substantial wall-clock speedup with virtually no cost to model intelligence or task-specific precision.
\section{Discussion}
\label{sec:discussion}

\begin{table}[t]
\centering
\caption{Comparison between \textbf{Structural Pruning} and \textbf{Quasar} for Qwen3. 
$\alpha$ represents the token acceptance rate, and $L$ denotes the mean acceptance length. 
Results highlight that training-free pruning fails to maintain the distributional alignment necessary for effective speculative decoding.}
\label{tab:pruning_failure}
\resizebox{\columnwidth}{!}{
\begin{tabular}{lccc}
\toprule
\textbf{Method} & \textbf{Retention / Precision} & $L$ & \textbf{Speedup} \\
\midrule
Vanilla (Full Model) & 100\% Layers / BF16 & 1.00 & 1.00$\times$ \\
\midrule
\rowcolor[gray]{0.9} \textit{Structural Pruning} & & & \\
~~Pruned-90\% & 90\% Layers / BF16 & 1.62 & 0.80$\times$ \\
~~Pruned-75\% & 75\% Layers / BF16 & 1.27 & 0.68$\times$ \\
~~Pruned-50\% & 50\% Layers / BF16 & 1.03 & 0.62$\times$ \\
\midrule
\rowcolor[gray]{0.9} \textit{Ours} & & & \\
~~\textbf{Quasar} & 100\% Layers / \textbf{W8A8} & \textbf{1.40} & \textbf{1.28$\times$} \\
\bottomrule
\end{tabular}
}
\end{table}

While Quasar achieves robust performance via quantized verification, we further investigate alternative strategies for verifier acceleration to justify our design choice. Specifically, we explore \textbf{Structural Pruning} (layer dropping) as a potential candidate for constructing a lightweight proxy verifier.

\textbf{The Failure of Training-Free Pruning.} 
A natural hypothesis is that the inherent depth redundancy in LLMs allows a structurally pruned model (e.g., retaining the first 75\% of layers) to serve as an efficient drafter. This approach theoretically avoids the complexities of hardware-level quantization while reducing the FLOPs required for the drafting phase.However, Table \ref{tab:pruning_failure} reveals a critical efficiency mismatch in training-free pruning. While conservative pruning (90\% and 75\% retention) maintains a relatively higher mean acceptance length ($L$), the computational overhead of these "heavy" drafters outweighs their predictive gains, resulting in a net slowdown (0.80$\times$ and 0.68$\times$). Conversely, while aggressive pruning (50\% retention) significantly reduces per-token latency, it induces a catastrophic distributional shift. The resulting acceptance length $L \approx 1.03$ indicates that the drafter's output is almost always rejected, rendering the speculative process futile.According to speculative decoding theory, a speedup is only achievable when the drafter strikes an optimal balance between low per-token latency and high alignment with the target model. The pruned verifiers fail this criteria: they are either too computationally expensive to provide a margin for speedup, or too inaccurate to sustain meaningful speculative gains. In contrast, Quasar achieves the necessary distributional alignment through quantization, delivering a 1.28$\times$ speedup by maintaining full model depth with minimal per-token cost.

\textbf{Why Quantization Wins.} 
This divergence offers a fundamental insight: \textit{Preserving the topological integrity of the network is more critical for verification than maintaining high numerical precision.} W8A8 quantization introduces uniform approximation noise across the entire parameter space but preserves the full depth and non-linear residual stream of the original model. In contrast, pruning removes entire functional transformations, creating an unrecoverable "feature gap" that leads to massive logit divergence. Consequently, for plug-and-play acceleration without expensive re-training or distillation, quantized verification emerges as the far more robust and effective paradigm.
\section{Conclusion and Future Work}
\label{sec:conclusion}

In this paper, we introduce \textbf{Quasar}, a novel framework that breaks the memory wall in speculative decoding by accelerating verification through quantized execution. While traditional speculative decoding research has focused primarily on reducing drafting costs, our empirical analysis reveals that the parallel verification stage has become a critical secondary bottleneck, constrained by the memory bandwidth requirements of high-precision model weights. Our core contribution leverages W8A8 quantization specifically for verifying draft tokens, preserving logit distribution fidelity while effectively halving memory traffic. Extensive experiments on models, such as OpenPangu and Qwen3, demonstrate an end-to-end throughput improvement of $1.28\times$ without compromising generation quality.

Despite its performance gains, Quasar's effectiveness is closely tied to the alignment between the quantized verification model and the original high-precision distribution. In complex reasoning tasks, even marginal quantization errors may lead to a slight decrease in the acceptance rate, potentially diminishing the speedup provided by reduced bandwidth. Furthermore, the current implementation relies on optimized INT8 kernels, which may limit portability across hardware platforms lacking specialized quantization support.

Looking ahead, several promising directions remain to be explored:
\begin{itemize}
    \item \textbf{Ultra-low Bit Verification:} Beyond W8A8, exploring lower-bit representations (e.g., 4-bit or 2-bit weight quantization) could further alleviate bandwidth pressure. We aim to investigate the precision threshold at which the degradation in verification accuracy outweighs the latency gains.
    \item \textbf{Dynamic Precision Scaling:} A potential extension involves dynamically adjusting verification precision based on the confidence of the draft model. For instance, employing high-precision verification for low-confidence tokens and low-bit execution for highly predictable sequences could further optimize the speed-accuracy trade-off.
    \item \textbf{Hardware-Aware Optimization:} Future iterations of Quasar will be tailored for specific hardware features, such as the INT8/INT4 tensor cores on modern GPUs or specialized NPU accelerators (e.g., Ascend 910C), to maximize computational throughput during the verification forward pass.
    \item \textbf{Integration with Tree-based Speculation:} We plan to evaluate the synergy of Quasar with more complex drafting mechanisms, such as tree-based speculation (e.g., EAGLE, Medusa), to determine its scalability in non-linear verification scenarios.
\end{itemize}

\nocite{langley00}

\bibliography{example_paper}
\bibliographystyle{icml2025}




\end{document}